# Beyond the molecular movie: dynamics of bands and bonds during a photo-induced phase transition

C. W. Nicholson[1,*,†], A. Lücke[2], W. G. Schmidt[2], M. Puppin[1], L. Rettig[1], R. Ernstorfer[1], M. Wolf[1,*]

[1]Department of Physical Chemistry, Fritz-Haber-Institut der Max Planck Gesellschaft, Faradayweg 4-6, 14195 Berlin, Germany.

[2]Department of Physics, University of Paderborn, Warburger Strasse 100, 33098 Paderborn, Germany.

[†] Present address: Department of Physics, University of Fribourg, Chemin du Musée 3, 1700 Fribourg, Switzerland

*Correspondence to:  christopher.nicholson@unifr.ch (C.W.N.), wolf@fhi-berlin.mpg.de (M.W.)

Ultrafast non-equilibrium dynamics offer a route to study the microscopic interactions that govern macroscopic behavior. In particular, photo-induced phase transitions (PIPTs) in solids provide a test case for how forces, and the resulting atomic motion along a reaction coordinate, originate from a non-equilibrium population of excited electronic states. Utilizing femtosecond photoemission we obtain access to the transient electronic structure during an ultrafast PIPT in a model system: indium nanowires on a silicon(111) surface. We uncover a detailed reaction pathway, allowing a direct comparison with the dynamics predicted by ab initio simulations. This further reveals the crucial role played by localized photo-holes in shaping the potential energy landscape, and enables a combined momentum and real space description of PIPTs, including the ultrafast formation of chemical bonds.

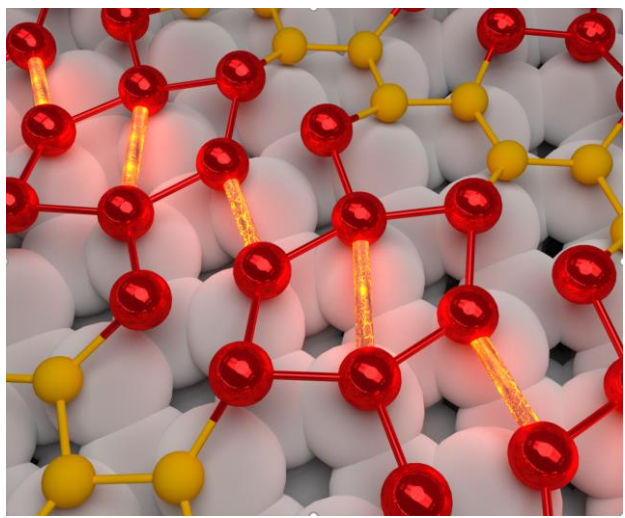

Artists view of the excitation and formation of chemical bonds along Indium nanowires (red balls) on a Silicon(111) surface during the ultrafast photoinduced phase transition between the 8x2 and 4x1 structures. This real space view of atoms and bonds is complemented by detailed measurememets of the electronic structure of electrons in their "momentum space" exhibiting the evolution of the band stuctrue providing a complete picture of the phase transition.

Reactive events in nature are associated with the formation or breaking of chemical bonds between atoms. Within the Born-Oppenheimer approximation (*1*) a description of reactions that separates the atomic and electronic degrees of freedom is employed, such that the atomic system evolves across a potential energy surface defined by the transient electronic structure. Testing these non-equilibrium concepts, whether in finite or extended systems, requires knowledge of both atomic and electronic structure on ultrafast time scales. Insulator-to-metal transitions allow a particularly stringent test of the relation between the two sub-systems, since the change in electronic structure is especially extreme, typically accompanied by a structural distortion. Ultrafast techniques have opened new avenues for exploring the interplay between the atomic and electronic sub-systems (*2–5*) including during photo-induced insulator-to-metal transitions (*6–8*), enabling the making of reciprocal space movies charting electronic structure dynamics (*7,9*), and "molecular movies" (*4,10*) which follow the real-time position of atoms during structural changes. Uniting these concepts to determine not only atomic positions but also the underlying electronic structure and the resulting reaction pathway along the potential energy surface (PES) has been a long sought after goal (*11*). Doing so requires access to the non-equilibrium electronic structure, for which time- and angle-resolved photoemission spectroscopy (trARPES) is ideally suited, as it allows direct access to the transient electronic states and their occupation in momentum space ($k$), including the interplay with the atomic lattice (*5,7,12*). The picture of electronic bands in periodic systems, often favored by physicists, is Fourier-equivalent to a real space ($r$) description of chemical bonds (*13,14*), which suggests the possibility of pursuing ultrafast bond dynamics in $r$-space (*15*) based on measurements in $k$-space (*16*). We realize this by determining the reaction pathway – including the full electronic structure dynamics – during an ultrafast structural phase transition at a surface, thereby going beyond the molecular movie concept.

A model phase transition system is found in atomic indium nanowires on the (111) surface of silicon – denoted In/Si(111) – which undergoes an order-order structural transition accompanied by an electronic insulator-to-metal transition (*17,18*). A close interplay between the electronic structure and specific lattice motions during the phase transition has been predicted which, in addition to a detailed knowledge of the equilibrium structure (*19–21*), makes this system ideal for investigating ultrafast changes in both $k$- and $r$-space. Recent time-resolved electron diffraction measurements have revealed that the structural PIPT completes within 1 ps (*22*) but such a technique does not give direct access to the underlying transient electronic dynamics.

Here we use trARPES to follow the ultrafast evolution of the electronic structure during the PIPT in In/Si(111) which, combined with *ab initio* molecular dynamics (AIMD) simulations, allows access to the microscopic forces driving the structural transition, and the full reaction pathway. In order to address the dynamic electronic structure of In/Si(111), we have developed a novel 500 kHz repetition rate extreme ultra violet (XUV) source at 22 eV (*23*), representing a significant advance compared with the state-of-the-art (*24,25*). This allows efficient access to the full, or even multiple Brillouin zones (BZ) in many materials. A schematic trARPES experiment is shown in Fig. 1A: the pump pulse (hν = 1.55 eV) excites electrons above the Fermi level ($E_F$) which are then ejected from the sample after a variable delay time $\Delta t$ by the probe pulse (hν = 22 eV) which have a cross correlation of 40 fs. In contrast to traditional ARPES (*26*), this allows simultaneous access to the electronic structure above and below $E_F$, illustrated for In/Si(111) in Fig. 1B.

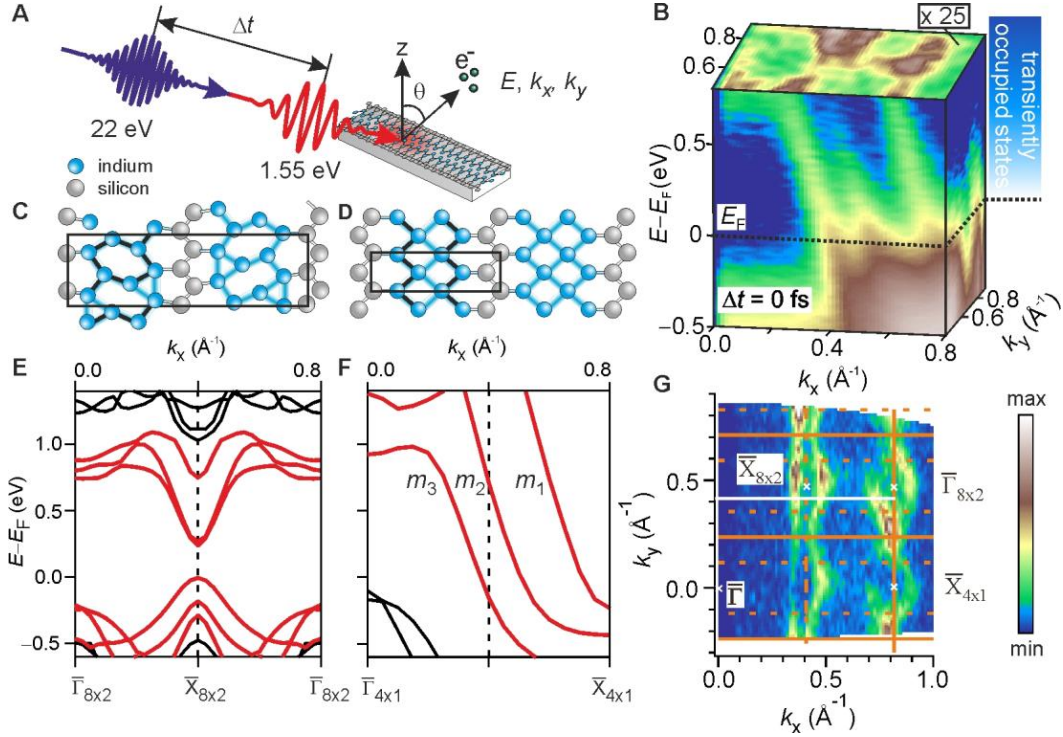

**Fig. 1. Experiment overview and material system.** (**A**) Schematic trARPES experiment, where $\Delta t$ is the variable delay between pump (red) and probe (purple) pulses. (**B**) Excited-state photoemission data (log-color scale) obtained at $T = 150$ K in the metallic (4x1) phase with an excitation fluence $F = 2$ mJ cm$^{-2}$. (**C**) Schematic $r$-space structure in the (8x2) phase and (**D**) in the (4x1) phase. Solid black lines highlight the structural motifs of the two phases, blue lines represent bonds. (**E**) Electronic band structure ($k$-space) calculated within the GW approximation in the (8x2) phase and in (**F**) for the (4x1) phase, corresponding to the structures in (C) and (D). The experimental characterization of the two phases is shown in the supporting online material (*23*). (**G**) Fermi surface obtained at 150 K with the 22 eV laser revealing the cut along which time-resolved measurements have been obtained (white line). Solid orange lines mark the (4x1) BZ boundaries, while dashed lines mark that of the (8x2) BZ. High symmetry points in the two phases are marked with crosses.

In/Si(111) undergoes a transition from an insulating (8x2) to a metallic (4x1) structure above 130 K (*27,28*), $r$-space schematics of which are shown in Fig. 1, C and D, respectively. The bonding motif in the insulating phase (Fig. 1C) consists of distorted hexagons, while in the conducting phase the In atoms rearrange into zig-zagging chains (Fig. 1 D). The $k$-space band structures of the two phases calculated within the *GW* approximation are given below the relevant structures in Fig. 1, E and F. In contrast to the (4x1) phase which has three metallic bands ($m_1$ - $m_3$) that cross $E_F$ (*17*), the (8x2) phase is gapped at the $\overline{\Gamma}_{8x2}$ and $\overline{X}_{8x2}$ points. Upon increasing the temperature across the (8x2) to (4x1) phase transition, the states initially lying far above $E_F$ at $\overline{\Gamma}_{8x2}$ shift down in energy and eventually cross $E_F$, forming the metallic $m_1$ band of the (4x1) phase. Concurrently the energy gap in the $m_2$ and $m_3$ bands at the $\overline{X}_{8x2}$ point closes, while at the same time the bands shift apart in momentum along the $k_x$ direction (*23*). We note that the three metallic bands predicted from the calculation in the (4x1) phase are clearly observed in Fig. 1B. The Fermi surface of the (4x1) phase in Fig. 1G shows the momentum cut along which our data are obtained.

To investigate the PIPT, the sample was cooled to 25 K and photo-excited by a pump pulse with incident fluence $F = 1.35$ mJ cm$^{-2}$, which corresponds to an excitation density in the surface In layer of around one electron per unit cell, implying homogeneous excitation. Selected snapshots following excitation are shown in Fig. 2, A to D. At $\Delta t = -450$ fs (Fig. 2A) the XUV pulse arrives before the pump pulse, hence the band structure reflects the unperturbed (8x2) phase with only states below $E_F$ occupied. Shortly after excitation at $\Delta t = 50$ fs (Fig. 2B) previously unoccupied states above $E_F$ are now clearly visible. An evolution of electronic states occurs, most clearly observed for the states around $\overline{\Gamma}_{8x2}$ ($k_x = 0.75$ Å$^{-1}$) which shift down in energy between $\Delta t = 50$ and $\Delta t = 250$ fs (Fig. 2C). At $\Delta t = 900$ fs (Fig. 2D) the system has fully transformed into the (4x1) phase. The overlaid *GW* band structure for the two phases highlights the occurrence of the PIPT.

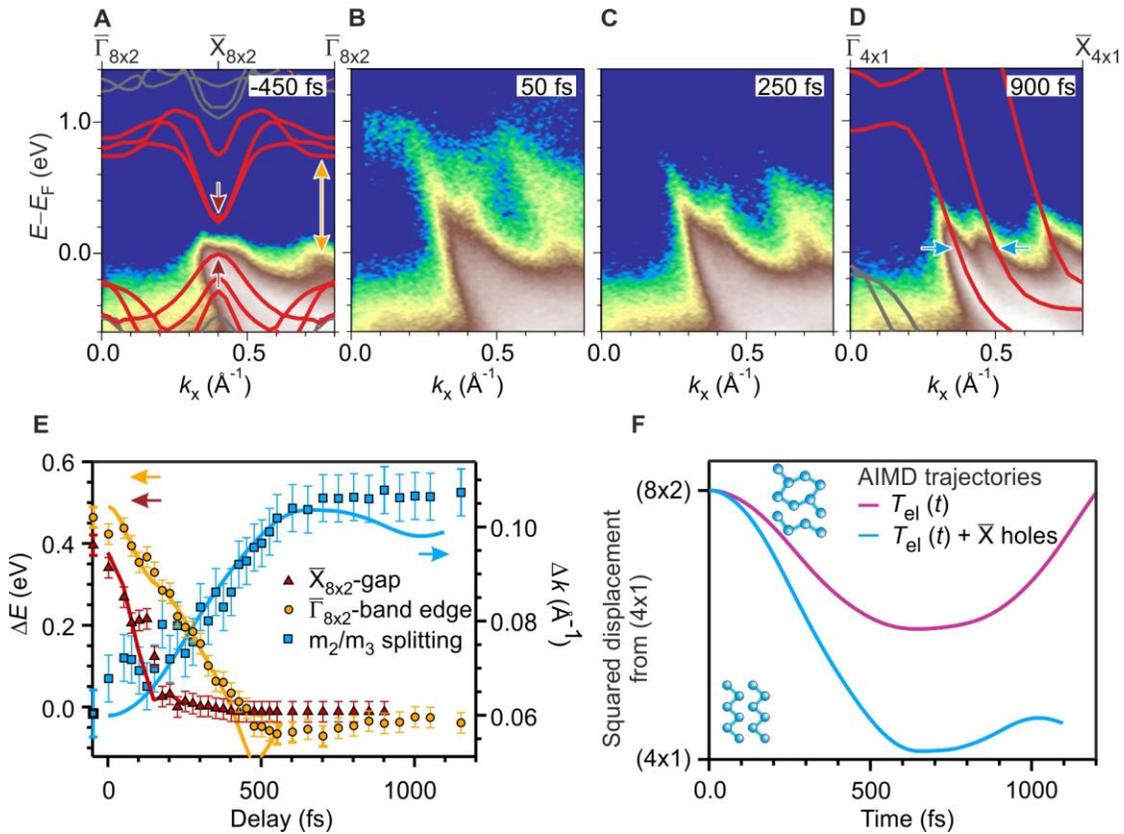

**Fig. 2. Electronic and atomic structure during photo-induced phase transition.** (**A**) to (**D**) trARPES data ($F = 1.35$ mJ cm$^{-2}$) on a logarithmic color scale at selected delays at a base temperature of $T = 25$ K. Arrows highlight the positions of the features of interest which are followed in (E). (**E**) Dynamics of the spectral regions marked by arrows in (A) and (D). Red data points track the size of the band gap at the zone boundary over time while the orange data mark the position of the band edge at the zone center with respect to the Fermi level. The blue data reveals the change of splitting between the two innermost bands marked in (D). Solid curves are the dynamics of the relevant spectral features from AIMD simulations, rescaled with respect to the *GW* band structure. For further details see the supporting online material (*23*). (**F**) Evolution of the atomic structure (AIMD trajectories) through the PIPT, showing the mean squared displacement of the atomic positions from the (4x1) phase following excitation: $\Sigma_i|R_i-R_{i,4x1}|^2$. Trajectories for two initial excitation conditions are shown: only for the distribution including

The dynamics of selected spectral features marked by arrows in Fig. 2, A and D chart the progress of the PIPT, and are plotted in Fig. 2E. The fastest dynamics are found at $\overline{X}_{8x2}$ (marked by the red arrow in Fig. 2A) where the band gap closes within 200 fs thus defining the ultrafast insulator-to-metal transition. As a second step, the conduction band edge at the BZ zone center (orange arrow) is found to reach $E_F$ after 500 fs. Finally the structural transition, as measured by the splitting between bands $m_2$, $m_3$ (Fig. 2D, blue arrow, and supporting online information), is completed after around 700 fs. This third time scale is in excellent agreement with the structural transition timescale observed by time-resolved electron diffraction which is completed after around 700 fs with a time constant $\tau = 350$ fs (*22*). It is remarkable that even before the structural transition is completed, two physically meaningful electronic transitions have occurred.

The distinct time scales of these three spectral features reveal a detailed pathway of the phase transition as it evolves along the electronic PES. To gain microscopic insight into the evolution of the atomic structure, electronic properties, and bond strengths along this pathway, we have performed AIMD simulations based on density functional theory (DFT) within the local density approximation (LDA), constrained by the experimental results. Since the experiment reveals the transient changes to the electronic states and their occupation, these can be used to simulate realistic excitation scenarios with AIMD. We have mapped the experimental *k*-space distribution of excited carriers across multiple BZs in Fig. 3. This reveals that electrons are strongly delocalized throughout the BZ, in contrast to photo-holes which are rather localized at the BZ boundary. Such a distribution is significantly different to the excitation conditions assumed in a previous study which forced excited electrons to be confined to the BZ center (*22*). In a first attempt, we assume transiently hot electronic distributions in the AIMD simulations based on the experimentally determined time-dependent electronic temperature (Fig. S6). As can be seen, however, the corresponding calculated trajectory (purple curve in Fig. 2F) describes an incomplete phase transition: the system starts to evolve from the (8x2) phase towards the (4x1) phase, but finally returns to the (8x2) ground state. This indicates that the LDA-DFT electronic structure is not sufficiently accurate. In fact, the inclusion of electronic self-energy effects within the *GW* approximation raises the energy of the uppermost zone boundary valence state by about 0.2 eV with respect to the zone center states (Fig. S7). Self-energy effects beyond the LDA thus lead to the preferential confinement of photo-holes at the BZ boundary as experimentally observed (Fig. 3). Unfortunately, AIMD simulations based on a self-energy corrected electronic structure are computationally prohibitively expensive. Therefore, we compensate the misalignment of the valence state energies on an ad hoc basis by fixing the occupation numbers (on top of the thermal occupation) in the AIMD simulations such that holes occur at the BZ boundary and the zone center valence states are occupied. The AIMD simulation based on this excitation scenario now indeed results in a complete phase transition (Fig. 2F, blue curve). This underlines the role of zone-boundary photo-holes as a key driving force in the structural transition. Moreover, the AIMD well reproduces all three time scales observed in the *k*-space experiment (Fig. 2E, solid lines), revealing an impressive level of accuracy in the simulated PES and the corresponding trajectory, even on these ultrafast time scales. The excellent agreement between our data and the simulations is strong evidence for the coherent directed motion of atoms within all unit cells during the PIPT, in agreement with the previous electron diffraction study (*22*). Such ultrafast directed dynamics cannot be explained by a statistical picture of the phase transition where different regions of the sample evolve incoherently (*23*).

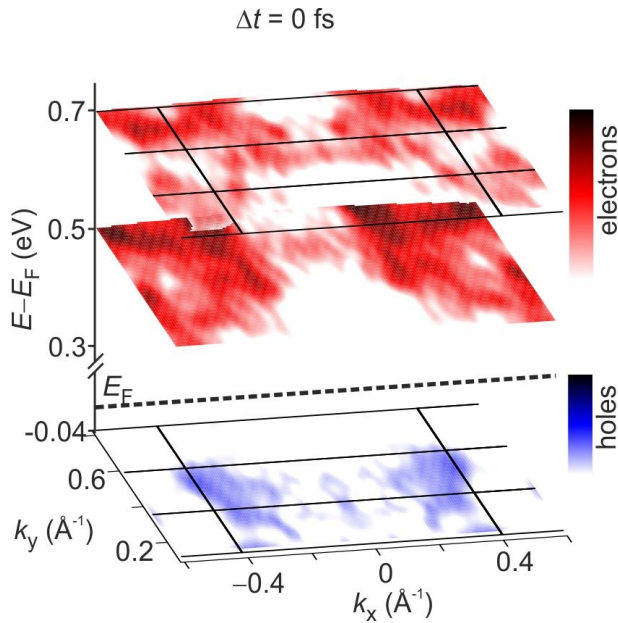

**Fig. 3. *k*-space distribution of excited carriers.** Experimentally measured difference map of the photoemission signal throughout multiple BZs in the (8x2) phase, revealing the distribution of excited electrons (red) and holes (blue) following photoexcitation ($F = 0.7$ mJ cm$^{-2}$). The distribution is obtained from the difference between spectra before excitation ($\Delta t = -1000$ fs) and $\Delta t = 0$ fs.

To further exemplify the high level of agreement between experiment and theory, in Fig. 4, A to C, the calculated band structure at three snapshots during the PIPT is compared with the corresponding band position at $\overline{\Gamma}_{8x2}$ extracted from our data. Both the calculated energetic position and the slope of the dispersion are observed to change in agreement with the experimental data. This exceptional agreement gives us the confidence to extract the *r*-space dynamics of nuclei and chemical bonds during the PIPT from the simulation. To do so we plot the electronic orbitals associated with the bands discussed above at the BZ center ($\overline{\Gamma}_{8x2}$) in Fig. 4, D to F, again for three snapshots. A transition from an orbital localized between opposite In hexagon atoms to a delocalized metallic state along the In chains is clearly seen during the PIPT.

To describe chemical bond formation additionally requires a measure of the bond strength. A quantitative understanding of bond strengths in extended systems can be gained from the crystal overlap Hamiltonian population (COHP) (*29,30*), which resolves each band into bonding and anti-bonding contributions as a function of energy; essentially a bonding character density of states for each electronic band. By performing a COHP analysis along the AIMD trajectory, we obtain the evolution of the surface bond strengths during the phase transition. Exemplarily, in Fig. 4, D to F, the formation of an In-In bond across the neighboring chains is shown. A gradual evolution of the bond strength up to 2 eV is observed, encoded in the blue to red color scale applied to the orbitals in Fig. 4, D to F. Combined with the orbital distribution this reveals the ultrafast formation of an In-In bond during the transition into the (4x1) structure, on the same time scale as the closing of the electronic gap in this region i.e. within 500 fs. The buildup of bond strength thus parallels the transition from a localized molecular orbital (insulator) to a delocalized (metallic) state during the phase transition.

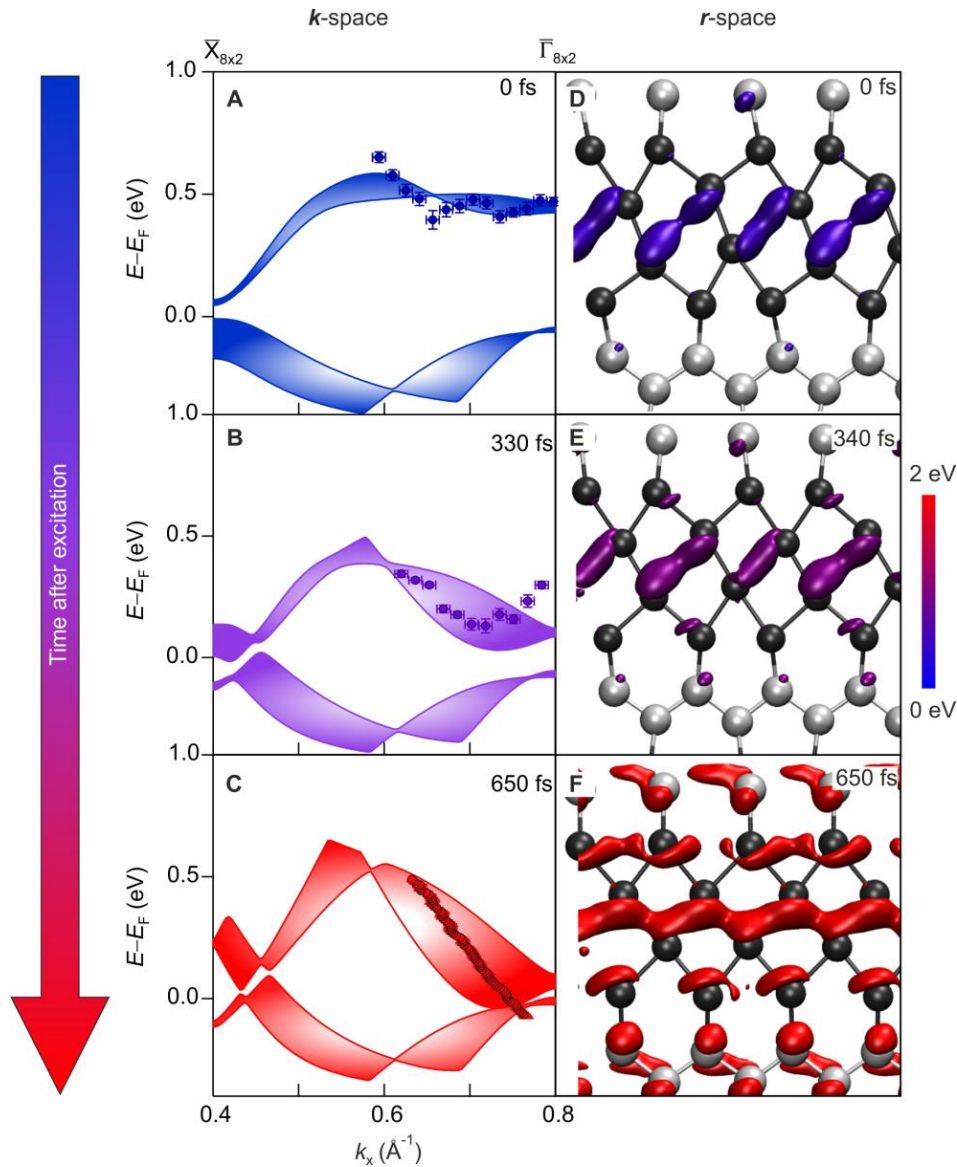

**Fig. 4. Dynamics of bands and bonds during the insulator to metal transition.** (**A**) to (**C**) Position of the *k*-space bands close to the $\overline{\Gamma}_{8x2}$ point at selected time delays extracted from the trARPES data, overlaid on the calculated LDA band structure (color filled for clarity). Errors bars mark a 95% confidence level. (**D**) to (**F**) Corresponding *r*-space dynamics of the orbital associated to the $\overline{\Gamma}_{8x2}$ band in (A) to (C). Both the shape of the orbital distribution and the bond strength – indicated by the color scale – change during the phase transition, as a bond across the indium hexagon is formed. A complementary picture of charge transfer during the bond formation and breaking, as well as movies of the full *k*- and *r*-space dynamics, may be found in the supporting online material (*23*).

From our analysis the following complete microscopic mechanism for the PIPT emerges: upon excitation, holes are created in the bonding states at $\overline{X}_{8x2}$, which correspond to In-In dimer bonds between the outer In chain atoms (*23*). Consequently, the dimer bonds characteristic for the hexagon structure weaken and break. At the same time, a significant fraction of excited electrons populates the states at $\overline{\Gamma}_{8x2}$ that are formed by a bonding combination of In states from neighboring In chains. Population of these excited states leads to interatomic forces that transform the hexagons into zig-zag chains leading to bond formation as shown in Fig. 4, D to F. The electron band related to these bonds ($m_1$) is lowered in energy as the In atoms contributing to this bond approach each other, further populating those states and strengthening the bond. It finally crosses the Fermi energy as shown in Fig. 4C resulting in the metallic state of the (4x1) phase.

Our combined experimental and theoretical approach extends the molecular movie concept by revealing the ultrafast electronic structure dynamics that govern a non-equilibrium structural transition. This unifying description bridges two fundamental concepts of physics and chemistry – band structure and chemical bonds – during ultrafast reactions. Besides elucidating the effect of the non-equilibrium electronic structure on structural dynamics, understanding the potential energy landscape induced following excitation paves the way for reaction pathways engineered via tailored excitation, potentially allowing optical control over such dynamic processes.

**Author Contributions :** C.W.N. prepared and characterized the samples. C.W.N. M.P. and L.R obtained the experimental data. C.W.N. analyzed the data. A.L. and W.G.S. performed the electronic structure calculations. W.G.S., R.E. and M.W. provided the project infrastructure. All authors discussed the results and their interpretation. C.W.N. wrote the manuscript with input and discussion from all authors. R.E. and M.W. were responsible for the overall project planning and direction.

**Acknowledgments:** We gratefully acknowledge funding from the Max-Planck-Gesellschaft, and the Deutsche Forschungsgemeinschaft through FOR1700, TRR142. The Paderborn Center for Parallel Computing (PC$^2$) and the Höchstleistungs-Rechenzentrum Stuttgart (HLRS) are acknowledged for grants of high-performance computer time. We thank Patrick Kirchmann, Claude Monney and Yunpei Deng for their contributions to developing the infrastructure of the trARPES experiment. The authors declare no competing interests. The data that underpin the findings of this study are available at
https://edmond.mpdl.mpg.de/imeji/collection/ph48BTv9YHGw49oc

# Supplementary Materials for

## Beyond the molecular movie: ultrafast dynamics of bands and bonds during a photo-induced phase transition

C. W. Nicholson, A. Lücke, W. G. Schmidt, M. Puppin,
L. Rettig, R. Ernstorfer, M. Wolf

Correspondence to: nicholson@fhi-berlin.mpg.de (C.W.N.), wolf@fhi-berlin.mpg.de (M.W.)

**This PDF file includes:**

> Materials and Methods
> Supplementary Text
> Figs. S1 to S8
> Author Contributions
> References (*1-47*)

**Other Supporting Online Material for this manuscript includes the following:**

> Movie S1
> Movie S2

**Materials and Methods**

Sample preparation and characterization

In/Si(111) nanowires were grown epitaxially on a Si(111) substrate miscut by 2° towards the [-1 -1 2] direction. The substrates (MaTecK GmbH) are *p*-doped with resistivity 0.075-0.085 $\Omega$ cm. Slightly more than 1 monolayer of pure In was evaporated from a home-built Knudsen cell at a rate of around 0.05 monolayers per minute onto a clean Si(111)-7x7 reconstructed surface at room temperature. The pressure during Si flashing never exceeded $1\times10^{-9}$ mbar, while during evaporation a typical pressure of $1\times10^{-10}$ mbar was maintained. Excess material was then removed by direct current annealing at around 500°C. The optimal coverage of 1 monolayer was judged live from the (4x1) pattern using low energy electron diffraction (LEED) shown in Fig. S1. A high sample surface quality is indicated by the sharp LEED spots and low inelastic background. A diffraction image of the (8x2) phase could not be obtained with the present setup as there is no option for sample cooling in the preparation chamber.

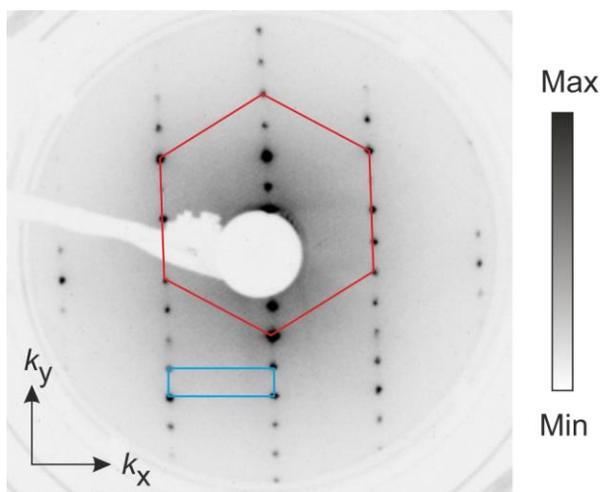

**Fig. S1. LEED characterization of In/Si(111)-4x1**
Low energy electron diffraction pattern of a freshly prepared In/Si(111)-4x1 phase obtained at 80 eV electron beam energy. The 4x1 (blue) and 1x1 (red) unit cells are overlaid. A single domain phase is obtained by using a substrate with a 2° miscut, thereby breaking the three-fold rotation of the (111) surface.

The static ARPES characterization of the sample presented in the main article in the two thermal phases is given in Fig. S2, obtained a He discharge lamp (21.2 eV). The (8x2) phase is reached by cooling the sample to 25 K with a liquid He cryostat. As can be seen in Fig. S2A the band structure is gapped and does not reach the Fermi level, in direct contrast to the (4x1) band structure that clearly shows the $m_1$, $m_2$, $m_3$ bands crossing $E_F$, shown in Fig. S2B. In the (8x2) phase the $m_2/m_3$ bands open a distinct band gap at around 0.4 Å$^{-1}$. In addition they coalesce into a single feature below $E_F$, in marked contrast to the well separated peaks observed in the (4x1) phase. This clear difference between the two phases reflects the different atomic arrangements of the underlying lattice. We therefore use the splitting between the two peaks to characterize the dynamics of the atomic structure during the ultrafast experiments (see below). In the $m_1$ band region it is no longer possible to follow a clear band dispersion in the (8x2) phase as the spectral weight is strongly reduced in this region due to the $m_1$ band being lifted above $E_F$. We are therefore able to spectroscopically distinguish between the two phases in our measurements. The weak residual intensity of the $m_1$ band is likely a result of small areas of the (4x1) phase that persist even at low temperatures which are pinned by defects or step edges, as commonly observed in the literature (*17,27,31–33*). Given the small spectral weight in this region it is clear that the area of the residual (4x1) phase is relatively small compared with the (8x2) phase. The dynamics we observe are therefore dominated by those of the (8x2) phase. We note that the apparent higher intensity of the $m_1$ band in the (8x2) phase in Fig. 2A of the main article is due to the logarithmic color scale chosen to highlight the weak population in the excited states.

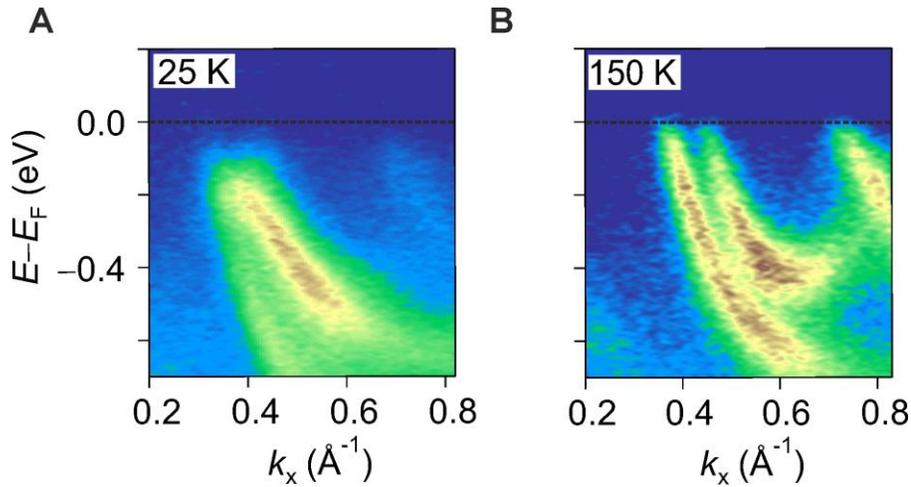

**Fig. S2. Characterization of In/Si(111) with ARPES**
ARPES spectra on a linear scale obtained with a He discharge lamp at 21.2 eV at 25 K (**A**) in the (8x2) phase and at 150 K (**B**) in the (4x1) phase. At low temperatures, a gap opens in the m2/m3 band region and the m1 band disappears above $E_F$ in contrast to the clear three band dispersion in the (4x1) phase. $E_F$ is marked by a dashed line in both cases.

Time- and angle-resolved photoemission spectroscopy

We employ a novel 500 kHz XUV light source at 22 eV produced by high-harmonic generation (HHG) in an Ar gas jet. The HHG is driven by the 1.55 eV output from a home-built OPCPA (*34*) with pulse energy of 30 µJ, which is then frequency doubled in a BBO crystal before entering the HHG chamber. A small focus spot (15 µm) and high backing pressure (2 - 4 bar) are used to obtain the necessary peak electric-fields and phase matching conditions for efficient HHG. A single harmonic is isolated by a combination of a multilayer XUV mirror, thin Sn filters, and a Si wafer at the Brewster angle to remove the driving frequency. A photon flux of $2\times10^{11}$ photons/s at the sample position is obtained at 22 eV. Time zero was determined using a bulk $WSe_2$ sample which shows a non-resonant two-photon response, allowing direct access to the temporal overlap of the two beams. The cross correlation (full width at half maximum) during typical measurements was between 35 fs and 40 fs. The overall energy resolution determined from the photoemission signal from a polycrystalline metallic sample was approximately 150 meV. An incident spot size of 120 x 95 µm$^2$ full width at half maximum was obtained for the probe beam, much smaller than the pump beam at 300 x 170 µm$^2$. The pump beam was linearly *s*-polarized and was incident at an angle of 15° to the sample surface normal. ARPES and trARPES measurements are obtained with a 2D hemispherical analyzer (SPECS GmbH) in conjunction with a 6-axis cryogenic manipulator which can be cooled to ~10 K (SPECS GmbH). Measurements were carried out at a base pressure of $2\times10^{-11}$ mbar.

Computational details

Density-functional theory (DFT) within the local-density approximation (LDA) in the Quantum Espresso implementation (*35*) is used to determine the atomic and electronic ground-state of In/Si(111). The surface is modelled using the supercell approach. The supercell contains three bilayers of silicon, the bottom layer of which is saturated with hydrogen. The electron-ion interaction is modelled with norm-conserving pseudopotentials. Plane waves up to an energy cutoff of 50 Ry are used to expand the electronic orbitals. The surface Brillouin zone is sampled using a 2x8 Monkhorst-Pack mesh. Quasiparticle band structures for the (8x2) and (4x1) surface phases (Figs. 1E, 1F, 2A, 2D, 4A-C) are obtained within the *GW* approximation (*36*), where the one-particle Green's function *G* and the screened Coulomb interaction *W* are obtained from the DFT electronic structure.

Ab initio molecular dynamics (AIMD) simulations of optically excited systems are performed within constrained DFT (*37*). The occupation of the electronic states is chosen in such a way that a specific excitation scenario is modelled, e.g., according to the Fermi-Dirac distribution for electronic temperature values obtained from the measured electron dynamics. At each time step in the simulation, the electronic structure is populated according to the thermal distribution obtained in the experiment, which is then used to calculate the forces on the atoms. The calculated atom dynamics then follow these forces until the next time step, which has a different population distribution, and so on. In this way the population changes are directly used to calculate the potential energy surface and the evolution of the system. The calculated trajectories are plotted in Fig. 2F and demonstrate the effect of the electronic excitation on the atomic system, and the timescale on which it evolves.

A crystal orbital Hamilton population analysis based on a projector augmented wave implementation (*38*) is used to quantify the surface bond strengths (Fig. 4D-F).

**Supplementary Text**

Analysis of the photo-induced phase transition

We have analyzed the photo-induced phase transition (PIPT) in terms of three spectroscopic signatures in distinct regions of $\boldsymbol{k}$-space – the zone boundary ($\overline{X}_{8x2}$), zone center ($\overline{\Gamma}_{8x2}$), and $m_2/m_3$ peak splitting – as detailed in the main text (Fig. 2). To exemplify our data analysis, data and fits from each of these regions are shown in Fig. S3.

For the analysis of the band positions and gap closing dynamics at the zone boundary and zone center, energy distribution curves (EDCs) are obtained at $k_x = 0.4$ Å$^{-1}$ and $k_x = 0.75$ Å$^{-1}$ respectively, as shown in Fig. S3A and C. The fit as well as the breakdown of the fit into individual components is shown for a single delay. At the zone boundary the EDCs are fitted with two Gaussian peaks plus a background in order to extract the position of both valence and conduction bands at each delay time during the pump-probe experiment; the difference, which corresponds to the band gap, is plotted in Fig. 2E. Data and fits at selected time delays are shown in Fig. S3B. The relevant dynamics at the zone boundary are those of the band edge at 0.4 eV, which corresponds to the $m_1$ band shifting towards the Fermi level, which are fitted with a Gaussian. The overall fits are shown together with the corresponding EDC at selected delay points in Fig. S3D. To highlight that the dynamics at the zone center are those of a shifting band, which cannot be explained by the redistribution of spectral weight alone, in Fig. S3E we have plotted the $k$-resolved position of the band at selected delay points. This reveals the gradual downwards shift of the whole $m_1$ band, similar to the data shown in Fig. 4 of the main article.

The simulated band gap dynamics have been extracted from the calculated time-dependent band structure (LDA). Since LDA is well known to underestimate band gap sizes, *GW* calculations have been carried out for the static (8x2) phase and the LDA structures have then been rescaled to match the *GW* results.

For the structural transition we followed the splitting of the $m_2/m_3$ peaks, obtained from momentum distribution curves (MDCs) at $E_F$, which are then fitted with two Lorentzian peaks plus a background (Fig. S3F). In Fig. S3G the position of the two peaks at each delay is overlaid on the corresponding MDC. The curve presented in Fig. 3E of the main text, revealing the structural phase transition dynamics, is obtained by subtracting the two peak positions shown in Fig. S3G. This particular feature of the spectral function was chosen as the occupied state bands are present at all times, even at long delays. The extracted dynamics match well with that found for the structural transition by the time-resolved electron diffraction study (*22*).

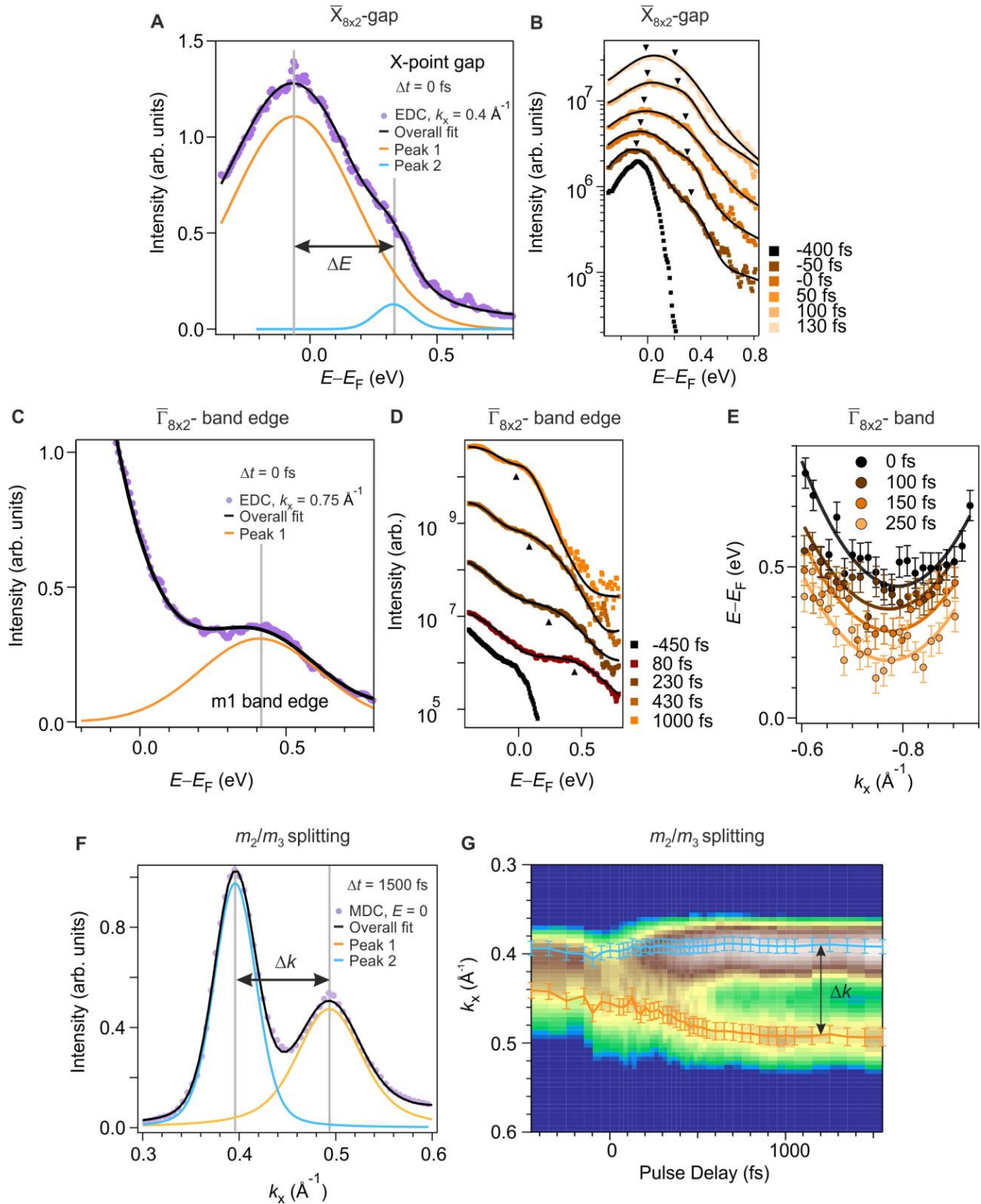

**Fig. S3. Analysis of the PIPT.**
EDCs of the trARPES data in the (8x2) phase at $\Delta t = 0$ fs at (**A**) the zone boundary and (**C**) at the zone center ($F = 1.35$ mJ cm$^{-2}$). The overall fits at selected time delays are shown in (**B**) and (**D**) respectively. Markers guide the eye. (**E**) k-resolved dispersion of the excited states at the zone center at selected time points, showing the evolution of the band to lower energies. Solid lines are guides to the eye. (**F**) MDC at $E_F$ at $\Delta t = 1500$ fs revealing the two peaks of the 4x1 phase. (**G**) Fit position for the two peaks in (**F**) overlaid on the respective MDC as a function of delay.

We exclude photovoltage shifts in the dynamics of the band structure, both for the In states close to $E_F$ and lower lying Si states, by measuring a pump-probe series of the full photoemission spectra including the secondary edge. In addition the time scale for surface photovoltage effects are known to be much longer than those that we observe in our measurements (*39,40*).

The data shown here are obtained in a relatively high fluence regime of close to one electron per unit cell, which implies a homogeneous excitation. This is obtained assuming a pump absorption of 0.5% in the In layer, consistent with Frigge *et al.* (*22*). Considering the long absorption length and predominance of indirect transitions at this wavelength in Si, the excitation density in the bulk Si substrate will be much lower than in the In surface layer. We resolve well defined band gaps and band dispersions in our data which suggests that the role played by disorder during the PIPT is negligible. Our analysis shows that we are able to follow in detail the evolution of specific electronic bands in energy, not only at specific momentum points but also momentum resolved (e.g. Fig. S3E). The fact that we can follow gradual changes to entire bands following excitation implies a coherent motion in which all unit cells behave in the same way. This is further corroborated by the comparison to the molecular dynamics simulations, which calculate the deterministic forces acting on all atoms in the unit cell to obtain their dynamics. The excellent agreement between our data and the simulations is therefore strong evidence of directed atomic motion, in which the PIPT evolves homogeneously via the motion of specific atoms within the unit cell, in agreement with previous diffraction studies of this quasi-one dimensional material (*22*) as well as analogous bulk materials (*41*). Such directed dynamics cannot be explained by a statistically driven phase transition where different regions of the sample evolve incoherently. The structural dynamics we observe are consistent with the quarter period of the combined frequencies of the atomic motions: 27 cm$^{-1}$ (1235 fs) and 17 cm$^{-1}$ (1962 fs) (*20*).

To further emphasize the deterministic nature of the PIPT, and that our AIMD simulations provide information also on the atomic dynamics, Fig. S4 shows the atomic speeds for both In and Si atoms following photo-excitation.

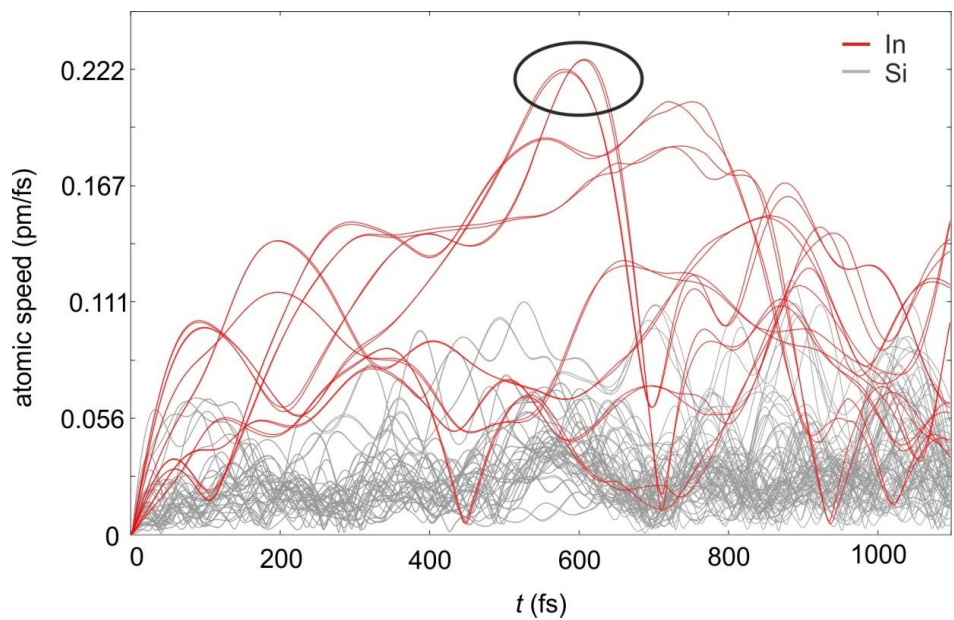

**Fig. S4. Atomic speeds during the photo-induced phase transition.**
Indium atom speeds are given by red curves, while silicon atoms are gray. The In atoms forming the bond shown in Fig. 4 are fastest (circled), about 0.22 pm/fs for a short time.

Distribution of spectral weight in ARPES

A common observation in ARPES measurements during phase transitions to symmetry broken ground states is that the spectral weight in the low-symmetry state has an intensity distribution reminiscent of the high-symmetry state (*42*). The amount of spectral weight transfer can be entirely accounted for by the electronic band gap that opens: In a mean-field model, the amount of spectral weight transfer is given by the ratio of bare and renormalized dispersions, where the renormalized dispersion is determined by the gap size. Such an approach has previously been used to explain spectral weight distributions in ARPES (*43*) and trARPES (*44*) during phase transitions. This is the reason that although the calculated band structure of the (8x2) phase is symmetric around the $\bar{X}$-point, the observed spectra is not. In order to show this in greater detail, we followed the procedure proposed in Ref. (*45*) and unfolded the band structure of the (8x2) surface into the Brillouin zone of the (4x1) unit cell as shown in Fig. S5A. In comparison with the band structure of the (4x1) structure Fig. S5B reveals that some (8x2) bands are indeed partially reminiscent of (4x1) bands, particularly in the region around the $\bar{X}$-point (bands $m_2/m_3$). However there are clear differences between the two phases (particularly at the Fermi level) that strongly deviate from the corresponding (8x2) states, as have been discussed in detail in the Materials and Methods section.

The deviation between the two structures is also clearly seen in the bands above $E_F$, as shown in Fig. S5, C and D. The data (identical to Fig. 2B, but with each MDC normalized) are obtained after excitation by the pump pulse at a delay of $\Delta t = 50$ fs, well before the phase transition occurs, and each horizontal row (MDC) is normalised to highlight the dispersion of the bands above $E_F$. The data are overlaid by the calculated band structure of the (8x2) phase in C and of the (4x1) phase in D. A particularly noticeable deviation between the two phases is found in the region of the $m_1$ band ($k > 0.5$ Å$^{-1}$). As has been described above, in the (8x2) phase this band is far above the Fermi level. The data agrees very well with the (8x2) band structure (Fig. S5C) and clearly deviates from the expectations of the (4x1) structure (Fig. S5D). This offers further proof that the sample is indeed in the (8x2) phase before excitation, and only transitions to the (4x1) phase due to photo-excitation. The presence of weak spectral weight following the $m_1$ band of the (4x1) phase is again visible, as discussed above.

An additional effect found in many photoemission experiments is that the distribution of spectral weight between Brillouin zones is not constant. This is a result of the photoemission matrix element, which places symmetry constraints on the electronic transitions into the final (measured) state. For this reason, the distribution in Fig. 3 is not purely periodic.

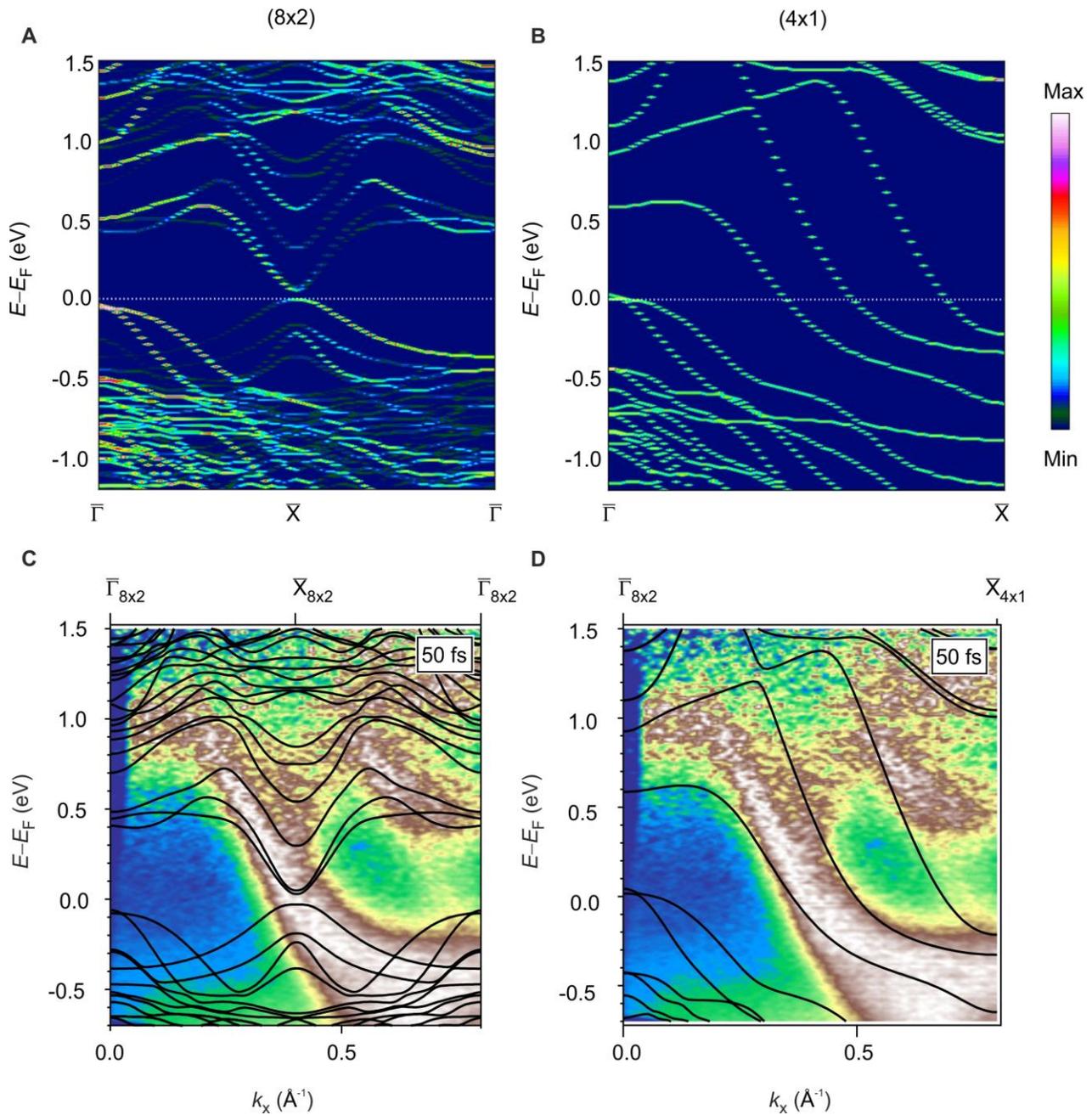

**Fig. S5. Distribution of spectral weight**
(**A**) Calculated LDA band structure of the (8x2) phase obtained by unfolding the electronic bands of the (8x2) phase into the primitive (4x1) cell according to the approach outlined in (*45*). The spectral weight is found to strongly favor the original band dispersion of the (4x1) phase (**B**), but clear differences in the band dispersion are maintained. (**C**) Normalized MDC images of the data presented in the main article (Fig. 2B) overlaid with the LDA calculated bands structure of the (8x2) phase. The position of the excited $m_1$ band above $E_F$ is well reproduced. (**D**) Same data as in (C), but overlaid with the calculated (4x1) structure, showing clear discrepancies in the $m_1$ band region ($k > 0.5$ Å$^{-1}$).

Analysis of the population dynamics: electronic temperature

We have extracted the transient electronic temperature during the PIPT (Fig. S6A) which has been used to constrain our ab initio molecular dynamics simulations (AIMD) (Fig. 2F). In order to assess whether a thermal description is indeed an appropriate description of the electronic population following excitation we compare the EDCs with their respective Fermi-Dirac function fits in Fig. S6B. EDCs are integrated over the region of the $m_2/m_3$ bands from 0.2 to 0.5 Å$^{-1}$, from which a transient temperature was extracted. The background signal is determined from the traces before excitation and is held constant during the subsequent delay fitting. A clear deviation from thermal behavior is observed during and shortly after excitation. However the deviation is small for $\Delta t = 50$ fs onwards. We therefore assume a thermal distribution to be an appropriate approximation. This time scale is typical for thermalization of the electronic system via electron-electron scattering (*46*). The electronic temperature curve is fitted with an exponential decay with $\tau = 285$ fs (Fig. S6A) which is then used directly as an input to the AIMD simulations to determine the constrained electronic population.

To corroborate this, we have evaluated the electronic energy density ($\varepsilon$) in the excited states by energy-weighting the population of states above $E_F$ as a function of delay. For a thermal system the electronic temperature is proportional to the square root of the electronic energy density, since $\varepsilon(t) = C_e T_e(t) = \gamma T_e^2$, where $C_e = \gamma T_e$ is the electronic heat capacity. Both the square root of the transient energy density and the electronic temperature are plotted in Fig. S6C. At early times during the pump pulse the fit tends to underestimate the temperature, evidenced by the fact that the electronic temperature consistently comes out lower than expected from the energy density. This agrees with the fits in Fig. S6B where the data shows a clear non-thermal component in addition to the thermal fit. The energy content analysis therefore reveals a thermalization time scale in agreement with that obtained from the EDC fit comparison (~50 fs).

The vibrational excitation of the lattice is difficult to address directly. While a simple two-temperature model of the system would indeed suggest a large value of the lattice temperature after 1 ps, this fails to account for the actual distribution of energy through the phonon system, which itself will not be thermalized (*47*). A large number of phonon modes exist in this system and are in fact efficiently excited (*22*). On which timescale the distribution of energy throughout all these modes thermalizes and leads to a non-equilibrium lattice temperature is non-trivial to address. However, an estimate of the lattice temperature at long delays and at excitation densities comparable with those that we present was obtained by an analysis of the diffuse scattered background (*22*). From this a maximum temperature increase of 30 K at a fluence of 3.1 mJ cm$^{-2}$ was obtained, which would keep the lattice well below the transition temperature of 130 K.

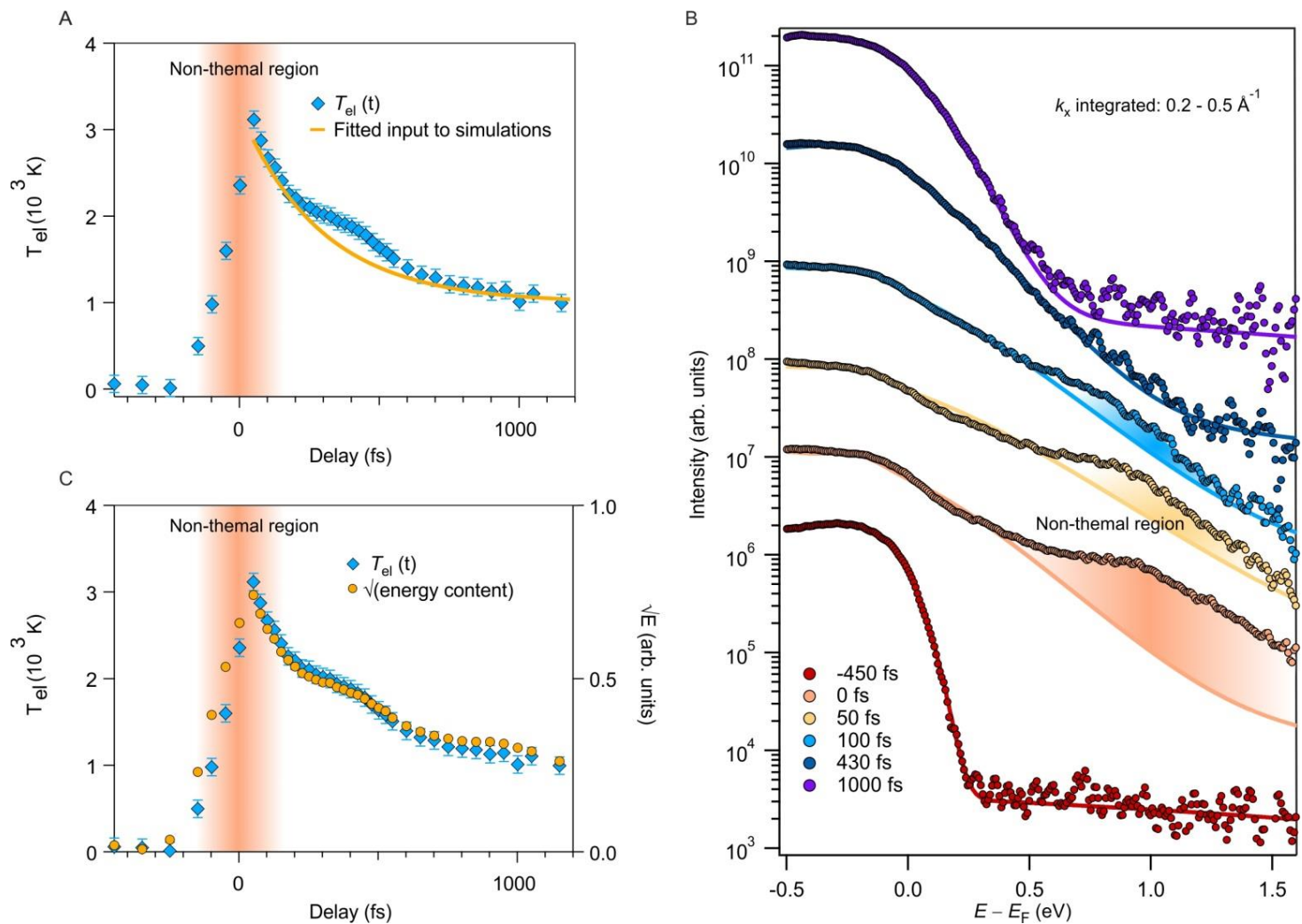

**Fig. S6. Thermal analysis of population dynamics.**
(**A**) Electronic temperature vs. pump probe delay from the region $k_x = 0.2 – 0.5$ Å$^{-1}$. The non-thermal region determined by the EDC analysis shown in (B) is shaded orange. An exponential fit is overlaid, which is used as an input to the molecular dynamics simulations. (**B**) EDCs at selected delays and their Fermi-Dirac fits, revealing non-thermal components at short delay times. (**C**) Square root of the energy content extracted from the data shown in (B) compared to the electronic temperature, corroborating the quasi-thermal description after excitation.

Comparison of LDA and GW band structures

The band structure of In/Si(111) in the (8x2) phase has been calculated within the local-density approximation (LDA) and GW approximation as described in the methods section above. A comparison of the two structures is presented in Fig. S7. Important differences are found at both the zone boundary and zone center. The LDA calculation at the zone boundary ($\bar{X}$) produces a gap in the electronic structure with an unrealistically small value (~60 meV) compared with that of the experiment (~350 meV at $\Delta t = 0$ fs). In contrast the GW value (~250 meV) is much closer to the experimentally obtained value. At the zone center ($\bar{\Gamma}$) LDA predicts the Si bands to be close to the Fermi level, only 30 meV below the In valence band states at $\bar{X}$. However in the more accurate GW calculation the Si states are shifted down by 220 meV to higher binding energies, while the In states remain close to $E_F$. The result is a band structure in which photo-excited holes preferentially sit at the zone boundary ($\bar{X}$-point) as is observed in Fig. 3 of the main text.

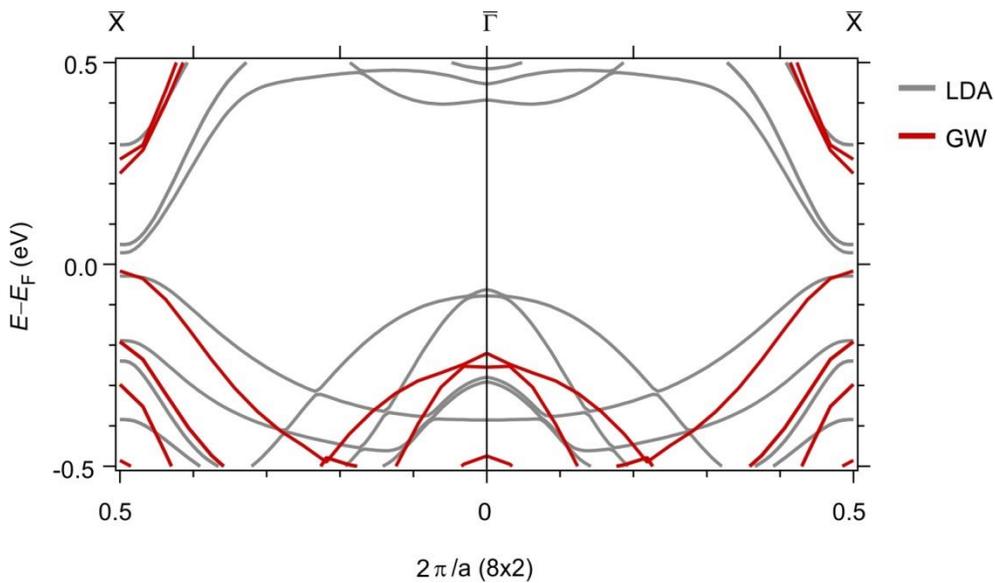

**Fig. S7. Comparison of LDA and GW band structures.**
Comparison of the In/Si(111)-8x2 phase within LDA (grey) and the GW approximation (red). The GW calculation is found to more closely reflect the experimental band gap at the $\bar{X}$-point. Additionally, the Si states at the $\bar{\Gamma}$-point are shifted down by ~200 meV with respect to the In states in the LDA calculation, explaining the observed preferential localization of holes at the $\bar{X}$-point.

Charge redistribution and bonding during the photo-induced phase transition

The photo-excitation redistributes electron density from In-In dimer bonds parallel to the In chain direction to states above the Fermi level, in particular to states ($m_1$ band) that correspond to In-In bonds across the In zigzag chains. This pulls the hexagon structure apart and leads to an atomic rearrangement, which in turn lowers the energy of the $m_1$ band towards the Fermi level. The corresponding bond across the In zigzag chains becomes more populated and is further stabilized. The *ab initio* calculations thus provide a very clear and detailed picture of the nature and driving force of the phase transition. In order to more clearly relate the momentum space information to a real space picture, Fig. S8 visualizes the excitation-induced charge transfer in real space; regions of electron loss (red) and gain (blue) directly relate to bond breaking and formation respectively of the bonds described above.

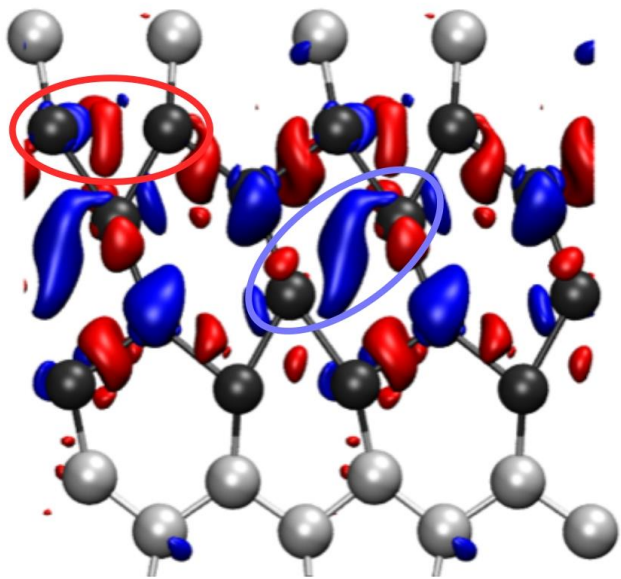

**Fig. S8. Charge redistribution and bonding during the PIPT.**
Calculated charge density difference between the (8x2) phase excited by the electronic distribution described in the main text (thermal distribution plus localized holes) and its electronic ground state. Electron accumulation is colored blue, while depletion is shown as red. A loss of charge/bonding strength at the outer In-In dimer bonds is observed (circled in red). The gain in the diagonal bond across the In hexagon (circled in blue) corresponds to that shown in Fig. 4 of the main text, which has bonding character, and stabilizes the photo-excited (4x1) phase.

Movies in momentum and real space

In Fig. 2 and Fig. 4 of the main text we have presented snapshots of the electronic structure and of the corresponding bond dynamics during the PIPT. In the included movies (S1 and S2) we show more frames relating to this data. In both cases the time delay $\Delta t$ is indicated.

**Movie S1. Photo-induced phase transition in momentum space.**
Photoemission intensity (linear scale) in energy and momentum as a function of pump-probe delay, showing the evolution from the (8x2) to (4x1) phases after photo-excitation at 0 fs.

**Movie S2. Ultrafast formation of a chemical bond in real space.**
Evolution of the atomic structure and electronic orbitals corresponding to the zone center bands ($m_1$) over time, revealing the ultrafast formation of a bond across the In hexagons. The color code represents the bond strength, as calculated from a COHP analysis. See main article for details.